\documentclass[pdflatex,sn-mathphys-num]{sn-jnl}

\usepackage{graphicx}%
\usepackage{multirow}%
\usepackage{amsmath,amssymb,amsfonts}%
\usepackage{amsthm}%
\usepackage{mathrsfs}%
\usepackage[title]{appendix}%
\usepackage{xcolor}%
\usepackage{textcomp}%
\usepackage{manyfoot}%
\usepackage{booktabs}%
\usepackage{algorithm}%
\usepackage{algorithmicx}%
\usepackage{algpseudocode}%
\usepackage{listings}%

\theoremstyle{thmstyleone}%
\theoremstyle{thmstyletwo}%

\theoremstyle{thmstylethree}%

\begin{document}

\title[Article Title]{Enhanced Interband Optical Nonlinearities from Coupled Quantum Wells}


\author*[1,2]{\fnm{Rithvik} \sur{Ramesh}}\email{rramesh@caltech.edu}

\author[1]{\fnm{Madeline} \sur{Brown}}\email{madeline.brown@utexas.edu}

\author[1]{\fnm{Amberly} \sur{Ricks}}\email{afr768@utexas.edu}

\author[3,4]{\fnm{Sedigheh} \sur{Esfahani}}\email{sesfahani@gradcenter.cuny.edu}

\author[1]{\fnm{Patrick} \sur{Devaney}}\email{patrick.devaney@utexas.edu}

\author[1]{\fnm{Kevin} \sur{Wen}}\email{kcwen11@utexas.edu}

\author[5]{\fnm{Moaz} \sur{Waqar}}\email{mwaqar@uci.edu}

\author[1]{\fnm{Zarko} \sur{Sakotic}}\email{zarko.sakotic@austin.utexas.edu}

\author[3]{\fnm{Sander A.} \sur{Mann}}\email{smann@gc.cuny.edu}

\author[1,11]{\fnm{Teddy} \sur{Hsieh}}\email{teddy\_hsieh@utexas.edu}

\author[1,10]{\fnm{Alec M.} \sur{Skipper}}\email{alecskipper@utexas.edu}

\author[1,7]{\fnm{Qian} \sur{Meng}}\email{qmeng19@utexas.edu}

\author[8]{\fnm{Hyunseung} \sur{Jung}}\email{hjung@sandia.gov}

\author[3]{\fnm{Michele}\sur{Cotrufo}}\email{mcotrufo@gc.cuny.edu}

\author[6]{\fnm{Farbod} \sur{Shafiei}}\email{farbod@physics.utexas.edu}

\author[6]{\fnm{Michael C.} \sur{Downer}}\email{downer@physics.utexas.edu}

\author[1]{\fnm{Sanjay} \sur{Shakkottai}}\email{sanjay.shakkottai@utexas.edu}

\author[7]{\fnm{Mark} \sur{Wistey}}\email{mwistey@txstate.edu}

\author[8]{\fnm{Igal} \sur{Brener}}\email{ibrener@sandia.gov}

\author[5]{\fnm{Xiaoqing} \sur{Pan}}\email{xiaoqinp@uci.edu}

\author[3,4]{\fnm{Andrea} \sur{Al\`u}}\email{aalu@gc.cuny.edu}

\author[1]{\fnm{Daniel} \sur{Wasserman}}\email{dw@utexas.edu}

\author[9]{\fnm{Jacob B.} \sur{Khurgin}}\email{jakek@jhu.edu}

\author*[1]{\fnm{Seth R.} \sur{Bank}}\email{sbank@utexas.edu}

\affil[1]{\orgdiv{Microelectronics Research Center and Electrical and Computer Engineering Department}, \orgname{The University of Texas at Austin}, \orgaddress{\street{110 Inner Campus Drive}, \city{Austin}, \state{TX} \postcode{78712}, \country{USA}}}

\affil[2]{\orgdiv{Electrical Engineering}, \orgname{California Institute of Technology}, \orgaddress{\street{1200 E California Blvd.}, \city{Pasadena}, \state{CA} \postcode{91125}, \country{USA}}}

\affil[3]{\orgdiv{Photonics Initiative, Advanced Science Research Center}, \orgname{City University of New York}, \orgaddress{\street{85 St. Nicholas Terrace}, \city{New York}, \state{NY} \postcode{10031}, \country{USA}}}

\affil[4]{\orgdiv{Physics Program, Graduate Center}, \orgname{City University of New York}, \orgaddress{\street{365 Fifth Avenue}, \city{New York}, \state{NY} \postcode{10016}, \country{USA}}}

\affil[5]{\orgdiv{Samueli School of Engineering}, \orgname{University of California, Irvine}, \orgaddress{\street{5200 Engineering Hall}, \city{Irvine}, \state{CA} \postcode{92697}, \country{USA}}}

\affil[6]{\orgdiv{Physics Department}, \orgname{The University of Texas at Austin}, \orgaddress{\street{110 Inner Campus Drive}, \city{Austin}, \state{TX} \postcode{78712}, \country{USA}}}

\affil[7]{\orgdiv{Physics Department}, \orgname{Texas State University}, \orgaddress{\street{601 University Dr.}, \city{San Marcos}, \state{TX} \postcode{78666}, \country{USA}}}

\affil[8]{\orgdiv{Center for Integrated Nanotechnologies}, \orgname{Sandia National Laboratories}, \orgaddress{\street{P.O. Box 5800}, \city{Albuquerque}, \state{NM} \postcode{87185}, \country{USA}}}

\affil[9]{\orgdiv{Department of Electrical and Computer Engineering}, \orgname{Johns Hopkins University}, \orgaddress{\street{3400 North Charles St.}, \city{Baltimore}, \state{MD} \postcode{21218}, \country{USA}}}

\affil[10]{\orgdiv{Department of Electrical and Computer Engineering}, \orgname{University of California Santa Barbara}, \orgaddress{\street{Institute for Energy Efficiency}, \city{Santa Barbara}, \state{CA} \postcode{93106}, \country{USA}}}

\affil[11]{\orgdiv{Electrical Engineering and Computer Science Department}, \orgname{Massachusetts Institute of Technology}, \orgaddress{\street{77 Massachusetts Ave.}, \city{Cambridge}, \state{MA} \postcode{02139}, \country{USA}}}


\abstract{The recent, rapid advances in nonlinear chipscale nanophotonics in the visible and near-infrared have been largely driven by manipulating the local dielectric environment proximate to decades-old workhorse bulk nonlinear optical materials, rather than increasing the inherent strength of their nonlinear response.  While proposed decades ago, we demonstrate the first experimental realization  of a new class of designer nonlinear materials that leverage the interband optical transition in asymmetric structures to provide strong second order susceptibility, $\chi^{(2)}$. Using simple AlGaAs/GaAs coupled quantum wells operating in the near-infrared as a prototype, we observed strong second harmonic generation enhancement of 1550 nm to 775 nm over bulk controls.  Extracted $\chi^{(2)}$ values were as high as 2750 pm/V, which is $>$7× that of bulk GaAs. Furthermore, measured susceptibilities agreed well with quantum mechanical calculations of  $\chi^{(2)}$ using layer profiles extracted from electron microscopy. Growth interruptions were employed to improve interfacial abruptness in response to electron microscopy characterization, resulting in increased  $\chi^{(2)}$ toward the simulation predictions for ideal heterointerfaces. More complex layer designs showed predicted  $\chi^{(2)}$ up to 7 nm/V. Such materials are anticipated to find myriad applications, including entangled photon generation at telecommunications wavelengths for chipscale quantum information processing.}




\maketitle

\section{Introduction}\label{sec1}

The importance and applications of optical nonlinearities have expanded tremendously since their first observation. At first, nonlinear optics relied on bulk crystals with large intrinsic nonlinear susceptibilities \cite{franken_generation_1961}. More recently with the development of chipscale photonics, there was a need for integrated nonlinear optical devices \cite{kong_super-broadband_2022, lu_ultralow-threshold_2021, wang_integrated_2018}. LiNbO$_{3}$ and many semiconductor alloys have become important material platforms for this purpose due to their relatively large bulk nonlinearities for applications including optical computing \cite{wetzstein_inference_2020, mcmahon_physics_2023}, quantum photonics \cite{chang_quantum_2014, wang_integrated_2020, moody_2022_2022}, chemical sensing and spectroscopy \cite{picque_frequency_2019, muraviev_massively_2018}, signal processing \cite{cotter_nonlinear_1999, langrock_all-optical_2006}, and terahertz sources \cite{owschimikow_resonant_2003, andronico_integrated_2008, jung_terahertz_2017}. 

Second order nonlinear optics are of particular interest for second harmonic generation, optical parametric amplification, and other processes, but require non-centrosymmetric materials \cite{boyd_nonlinear_2020, khurgin_origin_2015}. Furthermore, bulk nonlinearities are inherently weak, and many efforts to enhance them have relied on the use of metasurfaces for field concentration or polarization manipulation \cite{verhagen_enhanced_2007, krasnok_nonlinear_2018, lee_giant_2014}. Another promising approach to improve the strength of optical nonlinearities is designing materials with enhanced second order nonlinear susceptibility, $\chi^{(2)}$.   

The design of material structures with enhanced nonlinearities focuses on bandgap engineering to tailor the transition energies and transition dipole matrix elements between bound states. Large effective nonlinear susceptibilities can be achieved when the transition energies are near-resonant with the fundamental photon energy and when the matrix elements between states are large \cite{fejer_observation_1989}. The susceptibility, $\chi^{(2)}$, can be calculated from these quantum mechanical features of the material structure using the dipole matrix formalism \cite{bloembergen_light_1962, khurgin_secondorder_1987}. Various quantum well (QW) geometries were proposed to tune the dipole matrix elements and transition energies between bound states, including asymmetric coupled quantum wells \cite{khurgin_second-order_1988, sirtori_observation_1991}, stepped quantum wells \cite{boucaud_detailed_1990, scandolo_interband_1993}, and quantum wells with curved composition profiles \cite{bewley_far-infrared_1993}. The effect of asymmetric quantum well structures on electronic wavefunctions mimics, on a larger scale, heteropolar covalent bonds, which in bulk crystals tend to show large nonlinear susceptibilities compared to ionic bonds \cite{levine_bond-charge_1973, phillips_dielectric_1969}. Furthermore, designing heterostructure nonlinearities enables introducing asymmetry at the scale of the material layers. Since non-centrosymmetry is a requirement for second order optical nonlinearities, this structural asymmetry can enhance the nonlinearities of non-centrosymmetric crystals and can even enable nonlinearities in centrosymmetric, non-$\chi^{(2)}$ materials \cite{alloatti_second-order_2015}.

While many approaches have been explored, intersubband nonlinearities in multi-quantum wells have been the most successful. Enhancements of multiple orders of magnitude compared to the bulk nonlinearities have been demonstrated using resonant enhancement of intersubband nonlinearities in the conduction band. There have been many demonstrations of $\chi^{(2)}$ on the order of 10 nm/V in III-V QW structures (48 nm/V by Sirtori \textit{et al.} \cite{sirtori_observation_1991} and 28 nm/V by Fejer \textit{et al.} \cite{fejer_observation_1989}) for fundamental wavelength of $\lambda_f \approx 10 \, \mu$m, which is nearly 100-fold enhancement over the bulk susceptibility of GaAs (377 pm/V) \cite{boyd_nonlinear_2020, choy_accurate_1976, shoji_absolute_1997}. There have also been exciting results when coupling intersubband nonlinear enhancement with plasmonic metastructures, resulting in enhanced second order nonlinear processes as well as control over input and output polarizations \cite{lee_giant_2014, deng_giant_2020, yu_electrically_2022}. However, intersubband transition energies for $\chi^{(2)}$  are limited by the conduction band offset as depicted in Fig. \ref{fig: Concept}b.

Due to the success of intersubband nonlinear enhancement, interband nonlinear optics has remained under explored. In the late 1980’s, Khurgin presented theoretical predictions of $\chi^{(2)}$ enhancement using interband transitions in coupled asymmetric GaAs/AlGaAs QWs \cite{khurgin_secondorder_1987}. At the time, the predicted enhancement was only on the order of fractions of the bulk nonlinear susceptibility of GaAs \cite{khurgin_second-order_1988}. However, there is significant motivation to re-examine interband $\chi^{(2)}$, as leveraging transitions across bandgap energy opens the door to enhanced optical nonlinearities at technologically-relevant near-IR and visible wavelengths.

In this work, we have measured near-infrared second order susceptibilities of 1400 pm/V on average and 2750 pm/V at best from asymmetric coupled GaAs/AlGaAs quantum well structures, which is nearly seven times the already large susceptibility of bulk GaAs and ten times that of workhorse nonlinear materials such as LiNbO$_3$ \cite{boyd_nonlinear_2020}. The coupled QW designs were guided by quantum mechanical simulations and calculations of $\chi^{(2)}$ using the dipole matrix formalism, as shown in Fig. \ref{fig: Concept}c. The enhanced $\chi^{(2)}$ from second harmonic generation measurements are in good agreement with simulation predictions of $\chi^{(2)}$ based on material composition profiles from transmission electron microscopy, as depicted in Fig. \ref{fig: Concept}d. Lastly, design space exploration of the coupled QW structures predict achievable interband $\chi^{(2)}$ on the order of 7 nm/V, which represents over ten time enhancement of $\chi^{(2)}$ compared to bulk second-order nonlinear material platforms.

\section{Results}\label{sec2}

\subsection{Design of asymmetric coupled quantum well structures}\label{subsec2}

\begin{figure}
    \centering
    \includegraphics[width=1\linewidth]{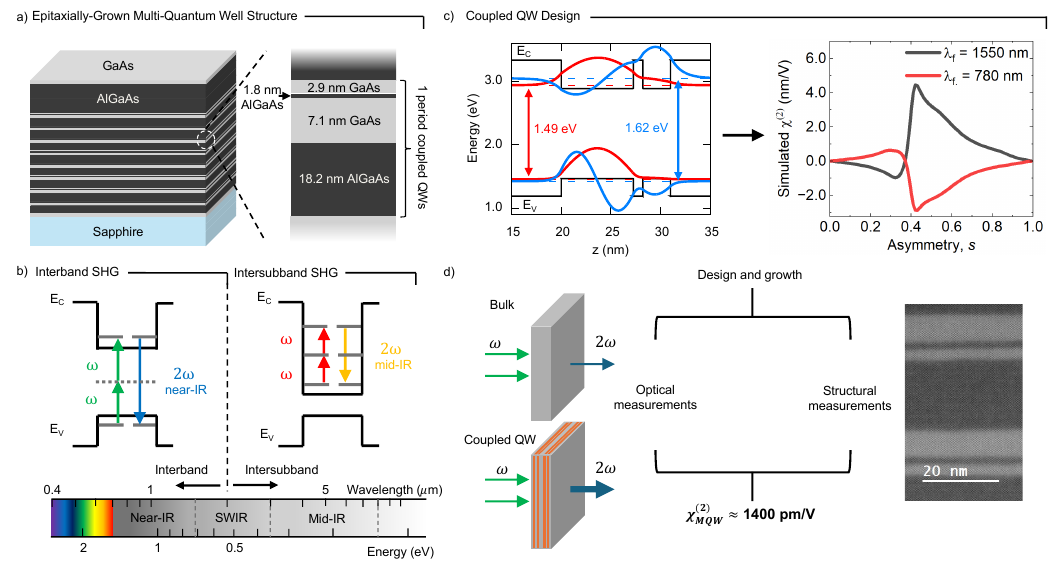}
    \caption{\textbf{Asymmetric coupled quantum wells for enhanced second order optical nonlinearity.} \textbf{a)} Diagram of the epitaxially-grown multi-QW samples with substrate transfer to sapphire. Each 30 nm period of coupled QW's is comprised of two GaAs QWs separated by a 1.8 nm AlGaAs tunneling barrier and an 18.2 nm AlGaAs period barrier between sets of coupled QWs. \textbf{b)} Interband second order nonlinearity utilizes transitions across the bandgap to access higher photon energies and shorter wavelengths compared to intersubband nonlinearities, which are limited by the achievable conduction band offset for a given material system. \textbf{c)} This band-edge diagram shows the coupled GaAs/Al$_{0.55}$Ga$_{0.45}$As QWs and the ground state (red) and first excited state (blue) energy levels and wavefunctions. The wavefunctions and energy levels are used to calculate $\chi^{(2)}$ as a function of QW asymmetry, which predicts the strongest $\chi^{(2)}$ for asymmetry \textit{s} = 0.42. \textbf{d)} This graphic shows the overall scope of the work. Quantum mechanical simulations guided the coupled QW design and sample growth. The average enhanced $\chi^{(2)} \approx$ 1400 pm/V from measured second harmonic generation was in good agreement with $\chi^{(2)}$ simulations of measured composition profiles from electron microscopy. Two sets of coupled QWs, as measured by STEM, are shown with GaAs layers in light gray and AlGaAs layers in dark gray.}
    \label{fig: Concept}
\end{figure}

The asymmetric coupled quantum well structure proposed in the 1980’s was revisited here in the context of interband nonlinearities \cite{khurgin_secondorder_1987, little_extremely_1987}. The difference in QW thicknesses introduces structural asymmetry, and the thin barrier induces wavefunction coupling between the two quantum wells. The QW asymmetry, \textit{s}, is defined as the ratio between the difference in the QW thicknesses and the sum of the thicknesses $(s=\frac{d_1 - d_2}{d_1 + d_2})$. The QW thicknesses, asymmetry, band offsets, and band gap were used to tailor the transition energies and dipole matrix elements to manually optimize $\chi^{(2)}$, as described in our prior publication \cite{ramesh_interband_2023}. The asymmetric coupled quantum well structure is depicted in Fig. \ref{fig: Concept}a. 

The GaAs/AlGaAs material system was chosen to target second harmonic generation in the near-infrared around 1550 nm, since the fundamental wavelength is not absorbed in the material. Furthermore, GaAs/AlGaAs is preferred for its well-known linear optical properties and relative ease of growth using molecular beam epitaxy (MBE). Both GaAs and AlGaAs are non-centrosymmetric crystals and have relatively large bulk nonlinear susceptibilities of 377 pm/V for GaAs and 290 pm/V for AlGaAs for near-IR wavelengths \cite{boyd_nonlinear_2020, choy_accurate_1976, shoji_absolute_1997, mobini_algaas_2022}. Here we show that this susceptibility can be surpassed through the asymmetric coupled quantum well structure.

We developed simulations of the coupled QW $\chi^{(2)}_{xzx}$, which henceforth we call $\chi^{(2)}$ for simplicity, using the density matrix formalism described in Methods and Supplemental Information \cite{khurgin_second-order_1988, khurgin_secondorder_1987, ramesh_interband_2023}. Through the simulations, we predict the structure $\chi^{(2)}$ of a few nm/V at near-IR fundamental wavelengths, which is an order of magnitude larger than the bulk second order susceptibility of GaAs. We predict that the structure $\chi^{(2)}$ is strongly dependent on the QW asymmetry, with the strongest susceptibility at \textit{s} = 0.42 as shown in Fig. \ref{fig: Concept}d. As expected, the structure $\chi^{(2)}$ goes to zero at both extremes of QW asymmetry, as they correspond to symmetric structures. The tunneling barrier thickness must be thick enough to split the QWs but not so thick that the wavefunction coupling is suppressed, and we predicted maximum $\chi^{(2)}$ for a $\approx$ 1 nm barrier thickness.

\subsection{GaAs/AlGaAs coupled QW structures exhibit enhanced second harmonic generation}\label{subsec2}

In this work, one period of asymmetric coupled quantum wells is defined as a set of two coupled QWs followed by 18 nm of AlGaAs period barrier, as shown in Fig. \ref{fig: Concept}a. Each period is 20 nm (10 nm total QW thickness, 1.8 nm barrier, and 18.2 nm period barrier). The samples were designed with multiple periods of coupled QWs, since the generated second harmonic intensity is known to scale with the square of the thickness of the nonlinear optical material in the absence of phase matching considerations for thin structures. 

The coupled QW samples were grown using MBE and transferred to transparent sapphire substrates to avoid the second harmonic background from a GaAs substrate. The second harmonic generation (SHG) response of each sample was measured using rotation-angle surface SHG in transmission at 1550 nm fundamental wavelength \cite{foster_study_2022, bloembergen_light_1962, shen_surface_1986}. Further sample growth and second harmonic measurement details are included in  Methods. 

\begin{figure}
    \centering
    \includegraphics[width=1.0\linewidth]{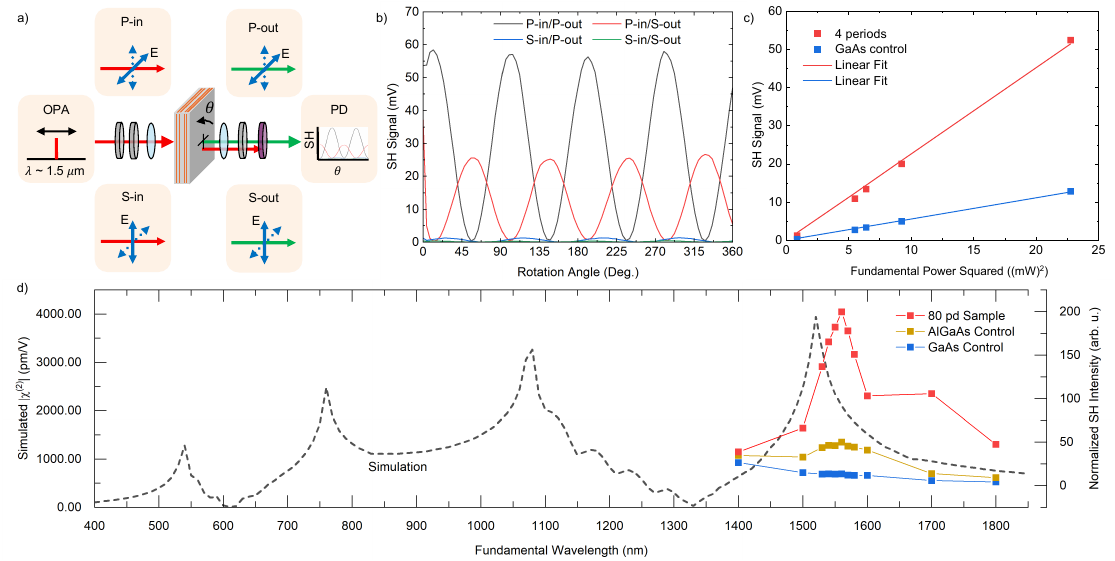}
    \caption{\textbf{Rotation-angle SHG measurements and signatures.} \textbf{a)} The sample is rotated through 360$^o$ of azimuthal angle (around the marked x on the sample face in the figure) while under constant illumination at the fundamental wavelength. The sample is tilted such that the polar angle is 45$^o$, ensuring a component of the P-in polarized electric field to be perpendicular to the QW plane. The input and output are p- or s- polarized, and low-pass filters block the fundamental wavelength while passing the SH wavelength. \textbf{b)} The rotational dependence of the SH signal shows a four-lobed periodicity, and the SH signal is strong for p-polarized input but weak for s-polarized input. \textbf{c)} The SHG power scaled with the square of the fundamental power, shown by a linear fit to the data plotted against the squared power axis, and the coupled QW sample (red) shows enhanced SHG compared to the bulk control (blue). \textbf{d)} The measured SHG dependence on fundamental wavelength shows a resonant peak for the QW sample, whereas there is no resonance behavior from the control. The measured wavelength dependence is in good agreement with the simulations for the coupled GaAs/Al$_{0.55}$Ga$_{0.45}$As coupled QWs with a 1.8 nm tunneling barrier, 10 nm total quantum well thickness, and QW asymmetry of $s=0.42$.}
    \label{fig: Setup and SHG Signature}
\end{figure}

The interband transition is excited when the fundamental field has a component perpendicular to the QW plane, which is accomplished for p-polarized input but not s-polarized. The rotation-angle surface SHG measurement shows four lobes due to the four-fold rotation-inversion symmetry of the crystal. An important characteristic of second harmonic generation is that the SH intensity scales with the square of the fundamental intensity. The polarization and rotation dependent SHG signatures are shown in Fig. \ref{fig: Setup and SHG Signature}b and the intensity dependence in Fig. \ref{fig: Setup and SHG Signature}c.

We measured and simulated the dependence of $\chi^{(2)}$ on the fundamental wavelength in Fig. \ref{fig: Setup and SHG Signature}d. The strongest $\chi^{(2)}$ occurs when the photon frequencies are close to resonant with the transition frequencies. For these asymmetric QW structures, the ground states in the conduction band and heavy hole band are separated by 1.49 eV (832 nm), and the first excited states are separated by 1.62 eV (765 nm). Thus, we expect peak $\chi^{(2)}$ when the SH photon energy is near-resonant with these transition energies. The simulated peaks in Fig. \ref{fig: Setup and SHG Signature}d around 760 nm and 1520 nm align with expected resonances with the transition between excited states. Using an optical parametric amplifier pumped by a Carbide (CB3-20W by Light Conversion) laser for wavelength tuning, we measured the SHG strength from the 80-period coupled QW structure for fundamental wavelengths between 1400 nm and 1800 nm. We observed a strong peak in SHG for wavelengths around 1560 nm, which was not present for the GaAs and AlGaAs control samples. The observed resonance closely aligns with the simulation predictions, and the lack of a resonance in the control samples indicates the effect of the coupled QW structures. The measured peak wavelength ($\approx$ 1560 nm) is slightly longer than the simulated peak ($\approx$ 1520 nm), which indicates a slightly lower transition energy in the MBE-grown sample compared to simulations. The strong agreement between the measured and simulated $\chi^{(2)}$ peak with fundamental wavelength demonstrates the effect of the QW structure on the optical nonlinearity. The pronounced resonance peak observed in SHG also suggests low optical losses, enabling high quality factor guided-mode resonance metasurfaces that will be reported in a future publication \cite{fathi_enhancing_2026}.

\subsection{Effective $\chi^{(2)}$ of coupled QWs}

Samples with 4, 12, and 16 periods of asymmetric coupled GaAs/AlGaAs QWs were grown and transferred to sapphire substrates. We observed enhanced second harmonic generation compared to the bulk GaAs control sample from all the coupled QW samples at 1550 nm fundamental wavelength. We expect that the effective $\chi^{(2)}$ of the multi-QW region is constant as the number of coupled QW periods is increased. For the same $\chi^{(2)}$, one would expect that the second harmonic power should scale with the square of the thickness of the multi-QW region (equivalently the number of periods) in the absence of phase-matching concerns. For the sub-wavelength thickness samples in this work, phase matching does not need to be considered. However, the many heterostructure interfaces result in standing waves within the sample, which break the scaling with sample thickness. Furthermore, the bulk AlGaAs layer, which was included to keep the linear optical properties consistent sample-to-sample, also produces significant second harmonic signal. For these reasons, the effective $\chi^{(2)}$ of the multi-QW region must be extracted from the measured SHG with consideration for standing-wave effects and bulk material contributions to SHG. 

The effective $\chi^{(2)}$ was found by building up the total second harmonic generation from each material layer weighted by electric field strength in each layer. The peak SH strengths from the rotation-angle SHG measurements of each sample were used to calculate $\chi^{(2)}$. The resulting effective $\chi^{(2)}$ accounting for standing wave effects and isolating the multi-QW contribution exhibited significant enhancement compared to the bulk GaAs susceptibility. The effective $\chi^{(2)}$ was around 1170 pm/V for the 12 and 16 period samples at 1550 nm fundamental, which is nearly quadruple the bulk $\chi^{(2)}$ of GaAs. The sample with 80 periods of coupled QWs exhibited $\chi^{(2)}$ of 1730 pm/V, and the sample with 4 periods of coupled QWs had a greater $\chi^{(2)}$ of 2750 pm/V. These results are shown in Fig. \ref{fig: STEM/EDS and simulation}c. The larger $\chi^{(2)}$ of the 4 period sample is likely due to the thin layer of QW's being positioned close to the interface with the sapphire substrate. Surface depletion effects can cause energy band-bending, which can change the transition energies resulting in resonant enhancement of interband $\chi^{(2)}$. 

\section{Discussion}\label{sec3}

\subsection{Structural measurements and effect on measured $\chi^{(2)}$}

\begin{figure}
    \centering
    \includegraphics[width=1\linewidth]{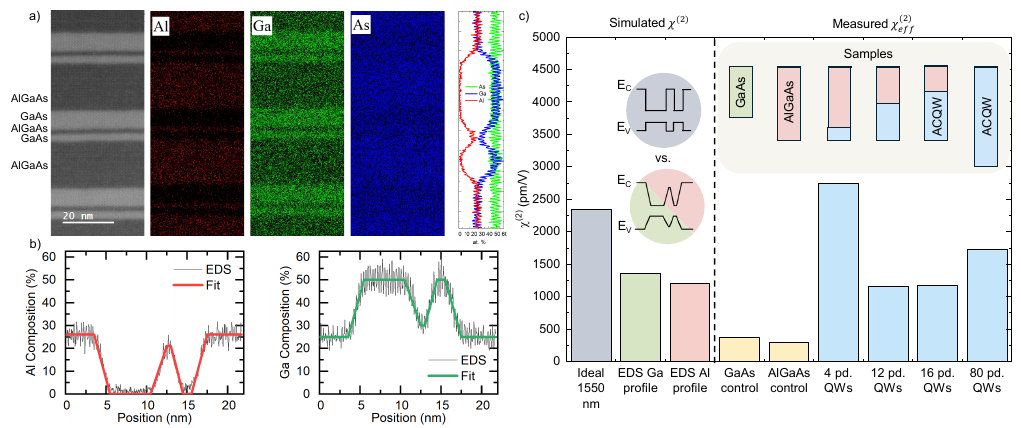}
    \caption{\textbf{STEM/EDS composition profiles and measured $\chi^{(2)}$}. \textbf{a)} Scanning transmission electron microscopy shows the asymmetric GaAs QW's coupled by the thin AlGaAs tunneling barrier. The energy dispersive spectroscopy (EDS) (colored Al, Ga, As, and profile plots) reveals the material composition profile of the layer structure. \textbf{b)} These plots show the measured EDS Ga and Al composition profiles, and the overlayed piecewise linear function is the composition profile to simulate $\chi^{(2)}$. \textbf{c)} This chart compares simulated $\chi^{(2)}$ at 1550 nm for the ideal coupled QW composition profile (gray), EDS composition profiles for Ga and Al (green and red), measured effective $\chi^{(2)}$ at 1550 nm for the GaAs and AlGaAs controls (yellow), and measured effective $\chi^{(2)}$ for the coupled QW samples (blue). The inset depicts each sample, their relative total thicknesses, and the positions of the multi-QW layers within the sample, oriented such that the sapphire substrate is at the bottom. The measured effective $\chi^{(2)}$ from the QW samples show enhancement compared to the controls, and they are in good agreement with the simulated $\chi^{(2)}$ based on the EDS composition profiles.}
    \label{fig: STEM/EDS and simulation}
\end{figure}

The measured $\chi^{(2)}$ was less than the simulation prediction of the ideal designed structure, which is seen comparing the simulated gray bar with the measured blue bars in Fig. \ref{fig: STEM/EDS and simulation}c. We investigated this discrepancy by characterizing the growth quality through scanning transmission electron microscopy (STEM) and energy dispersive spectroscopy (EDS). We measured the 16-period sample using STEM and EDS to determine the actual thicknesses and alloy compositions of each layer. We found significant gradients in aluminum composition across interfaces between GaAs and AlGaAs, as shown in the EDS measurements of Fig. \ref{fig: STEM/EDS and simulation}a. The gradients were mostly linear across 1 nm. 

To understand the effect of the compositional gradients, we simulated $\chi^{(2)}$ with the measured composition profiles. Separately simulating $\chi^{(2)}$ for 1550 nm fundamental according to the Al and Ga profiles (Fig. \ref{fig: STEM/EDS and simulation}b) resulted in predicted $\chi^{(2)} = 1200$ pm/V and $\chi^{(2)} = 1363$ pm/V, respectively. This is shown by the red and green bars in Fig. \ref{fig: STEM/EDS and simulation}c. Thus, the simulated $\chi^{(2)}$ from the real, non-ideal structure is in agreement with the measured effective $\chi^{(2)}$ for 1550 nm fundamental wavelength. The $\chi^{(2)}$ simulation of the EDS-based composition profile also demonstrates the sensitivity of the structure to thickness variations on the order of a nanometer. Such changes affect the energy levels (and thus transition energies) as well as the bound state wavefunctions, thus changing the resonance condition at a given photon energy and changing the transition dipole matrix elements. Overall, the STEM/EDS results and the simulation analysis suggest that improving the growth procedure to achieve sharper interfaces can enable greater enhancement in $\chi^{(2)}$. 

\begin{figure}
    \centering
    \includegraphics[width=1\linewidth]{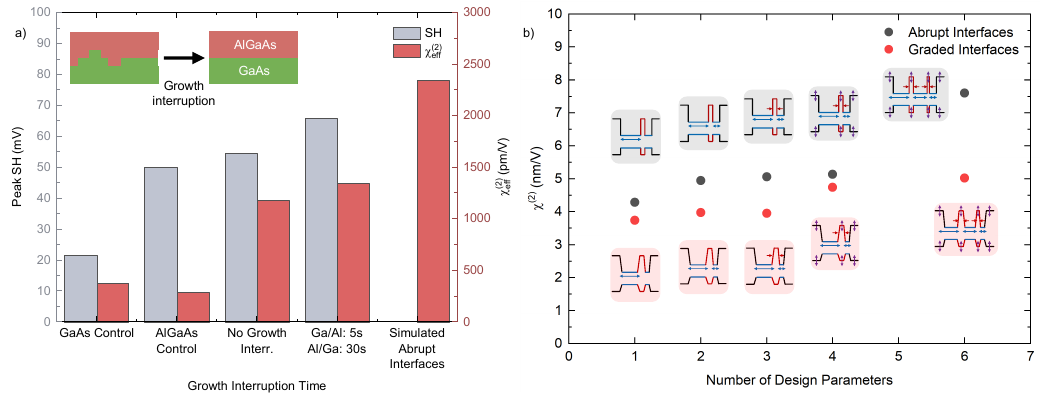}
    \caption{\textbf{Improved SHG by growth interruptions and large scope for designing tailored nonlinearities}. \textbf{a)} Growth interruptions at the GaAs/AlGaAs interfaces improve SHG (gray) and the effective $\chi^{(2)}$ for the 16-period coupled-QW structures at 1550 nm fundamental wavelength. \textbf{c)} An enhanced $\chi^{(2)}$ up to 7 nm/V was predicted by simultaneously optimizing the number of coupled QWs and the thicknesses of the QWs and tunneling barriers. The simulated data points and inset diagrams in gray use ideal interfaces, whereas the data points and insets in red use graded interfaces as measured by STEM. Even with graded interfaces, it is possible to significantly enhance interband $\chi^{(2)}$ with more complex coupled QW designs.}
    \label{fig: Design Space}
\end{figure}

Interrupting the MBE growth at the heterointerfaces resulted in improved second harmonic generation, as shown in Fig. \ref{fig: Design Space}a. Surface scattering, transport, and photoluminescence measurements support that growth interruptions improve surface smoothness at the interfaces, which can improve the observed composition gradients \cite{sakaki_interface_1987, neave_dynamics_1983, herman_heterointerfaces_1991}. Pausing for 5 seconds at the GaAs to AlGaAs interfaces and 30 seconds at the AlGaAs to GaAs interfaces resulted in an enhanced $\chi^{(2)}_{eff} \approx$ 1345 pm/V, which is an improvement toward the simulation prediction of 2340 pm/V based on abrupt composition changes at the interfaces. Further exploration of growth interruption and interface smoothening techniques may potentially unlock even greater $\chi^{(2)}$ from the coupled QW structures.

\subsection{Physical origin of QW $\chi^{(2)}$}
Analyzing the dipole matrix formulation of the coupled QW $\chi^{(2)}$ reveals that the dominant physical effect driving the interband $\chi^{(2)}$ enhancement is from the diagonal intersubband matrix elements. This is supported by simulations showing pairwise cancellations of the contributions to $\chi^{(2)}$ from the possible three-wave mixing transitions except the transitions with diagonal intersubband matrix elements \cite{ramesh_interband_2023}. Furthermore, the dominant diagonal intersubband matrix elements resemble the dipole moments that give rise to bulk optical nonlinearities. In this perspective, we expect that the second harmonic intensity scales with the square of the fundamental intensity, which agrees with our observations in Fig. \ref{fig: Setup and SHG Signature}. The analysis of the dipole matrix formalism for $\chi^{(2)}$ is provided in supplemental information.

\section{Conclusion}
Here we have demonstrated bandgap-engineered coupled QW structures for enhanced interband $\chi^{(2)}$, which represents a significant step towards designer nonlinear materials at technologically-important near-infrared wavelengths. The observed polarization dependence and resonant wavelength indicate the QW nonlinearity. Furthermore, the measured enhancement of SHG and corresponding enhanced $\chi^{(2)} \approx$ 1400 pm/V are in good agreement with $\chi^{(2)}$ calculated from measured composition profiles. The improvement in SHG from growth interruptions further solidifies the connection between QW structure and $\chi^{(2)}$ enhancement. 

By simultaneously tailoring both structure and material parameters, there is huge scope for designing the enhancement in nonlinearity for any application requirements. Alloy compositions and III-V material choice can be used to precisely tailor the wavelength for resonant enhancement of interband nonlinearities. By further explorations of the design space, we predict coupled QW designs with up to $\chi^{(2)} \approx$ 7 nm/V, as shown in Fig. \ref{fig: Design Space}b. While we focused here only on the established AlGaAs/GaAs materials system, designs can be extended to shorter wavelengths using wider bandgap semiconductors or more complicated semiconductor heterostructures, such as digital alloys and type-II superlattices. Digital alloys possess a vast design space to finely tune carrier confinement potentials, transition energies between states, and electronic wavefunctions \cite{maddox_broadly_2016}. Broadening of the design space provides further opportunity to simultaneously design for multiple properties, such as simultaneously high $\chi^{(2)}$ and low linear loss, $\alpha(2\omega)$. This capability could be particularly enabling for quantum photonic operations in quantum interconnects where losses are particularly problematic, including significantly improving entangled-pair source brightness, reducing squeezing threshold power, and increasing the efficiency of quantum frequency converters and nonlinear Bell-state analyzers \cite{helt_how_2012}. There is also scope to use metasurfaces in conjunction with the interband enhancement structures for plasmonic field enhancement \cite{krasnok_nonlinear_2018, lee_giant_2014}. The field enhancement can drive significant increases in nonlinear response, and the metasurfaces can be designed to selectively excite certain $\chi^{(2)}$ tensor elements.

\section{Methods}\label{sec11}

\subsection{$\chi^{(2)}$ calculation and quantum mechanical simulations.}

The general form of the dipole matrix formalism equation for $\chi^{(2)}$ is \cite{boyd_nonlinear_2020, bloembergen_nonlinear_1996, khurgin_secondorder_1987}

\begin{multline}
    \chi_{ijk}^{(2)}(\omega_1,\omega_2) = \frac{N_ze^3}{2\epsilon_0\hbar^2}\sum_P\sum_{b_1, b_2, b_3}\sum_{l,m,n}f_{b,l}\sum_{k_{||}} \\\frac{\langle\phi_{1,l}^*|r_i|\phi_{2,m}\rangle\langle\phi_{2,m}^*|r_j|\phi_{3,n}\rangle\langle\phi_{3,n}^*|r_k|\phi_{1,l}\rangle}{[\omega_{b_2,m}(k_{||})-\omega_{b_1,l}(k_{||})-\omega_1-\omega_2][\omega_{b_3,n}(k_{||})-\omega_{b_1,l}(k_{||})-\omega_2]}
    \label{eqn: chi(2) starting equation}
\end{multline}

Where $\phi$ are the wavefunctions of the denoted state and band, $\omega_1$ and $\omega_2$ are the input frequencies, $\omega_{b,(m,n,l)}(k_{||})$ is the frequency associated with the energy of the indexed state and band at the given in-plane $k$ state, and the summations are over the polarization combinations ($P$), bands ($b_1,b_2,b_3$), states ($m, n, l$), and in-plane $k$ states. 

The unit cell and envelope portions of the wavefunctions were separated using the Bloch formalism. The resulting interband and intersubband matrix elements in the numerators only use the envelope portions of the wavefunctions, and the interband matrix element of the unit cell wavefunctions ($r_{e,hh}$ in Eq. \ref{eqn: chi(2) equation}) was pulled out of the summation. Since the ground state of the second harmonic transitions are in the filled valence band, the fermi function, $f_{b,l}$, is unity. The first two bound states in the heavy hole and conduction bands are the only states considered because we were guaranteed to have two bound states in each band across the range of structures simulated in this work. Applying these simplifications for the coupled QW structures, the $\chi^{(2)}$ equation was simplified to Eq. \ref{eqn: chi(2) equation}. To calculate $\chi^{(2)}$ using Eq. \ref{eqn: chi(2) equation}, the summation over in-plane k states was converted to an integral over $(k_x, k_y)$. We found the contribution to the $\chi^{(2)}$ from in-plane k states saturated by one-tenth of the Brillouin zone away from zone center, so we integrated from zone center ($k_{||} = 0)$) to one-tenth of the Brillouin zone \cite{ramesh_interband_2023}.

\begin{multline}
            \chi_{xzx}^{(2)}(\omega_1,\omega_2) = \frac{N_z e^3 r_{e,hh}^2}{6\epsilon_0 \hbar^2}\\\sum_{k_{||}}\sum_{m,n}\sum_{l}\Bigl(\frac{\langle\psi_{hh,m}|\psi_{e,n}\rangle\langle\psi_{e,n}|z|\psi_{e,l}\rangle\langle\psi_{e,l}|\psi_{hh,m}\rangle}{(\omega_{hh,m}^{e,n}(k_{||})-\omega_1-\omega_2+i\Gamma)(\omega_{hh,m}^{e,l}(k_{||})-\omega_1+i\Gamma)} - \\\frac{\langle\psi_{e,n}|\psi_{hh,m}\rangle\langle\psi_{hh,m}|z|\psi_{hh,l}\rangle\langle\psi_{hh,l}|\psi_{e,n}\rangle}{(\omega_{hh,m}^{e,n}(k_{||})-\omega_1-\omega_2+i\Gamma)(\omega_{hh,l}^{e,n}(k_{||})-\omega_1+i\Gamma)}\Bigl)
    \label{eqn: chi(2) equation}
\end{multline}

In eq. \ref{eqn: chi(2) equation}, [m,n,l] index the bound states in the conduction and heavy hole bands,  $\psi_{e,hh;m,n,l}$ are the envelope wavefunctions for the indexed band and state, $\omega_{1,2}$ are the input photon frequencies, $\omega_{hh;m,l}^{e;n,l}$ are the frequencies associated with the interband transitions between the indexed states, $\Gamma$ is the line broadening and is assumed to be 5 meV, $k_{(||)}$ are the in-plane k states, $N_z$ is the number of quantum wells per unit length, and $r_{(e,hh)}=\langle u_e^*|r|u_hh \rangle$ is the interband matrix element of the unit cell portions of the wavefunctions in the conduction and heavy hole bands.

The envelope wavefunctions for the asymmetric coupled QW structures were determined using Schrodinger-Poisson methods with the Nextnano software \cite{birner_nextnano_2007}. The interband matrix element of the unit cell wavefunctions of GaAs was determined using density functional theory with the Vienna ab initio Simulation Package using HSE06 hybrid orbitals. 



\subsection{Sample preparation: molecular beam epitaxy and substrate transfer.}

The samples (designed layer structures) in this study were grown on semi-insulating GaAs (100) wafers using molecular beam epitaxy (MBE) in a Varian Gen II system. The system is equipped with a solid-source valved cracker for As and solid-source thermal effusion cells for Al and Ga. The substrate temperature was measured by band edge thermometry. The materials were grown under a 15x As overpressure. The AlGaAs growth rate was 1.85 \AA/s, the GaAs growth rate was 0.83 \AA/s, and the growth temperature was 600$^o$ C. Growth interruptions were introduced by stopping material growth by closing the Al and Ga shutters. By maintaining an ambient As overpressure, the surface of the sample can smooth out.

Because the bulk GaAs substrate produced a large second harmonic background in the measurements, it was necessary transfer the asymmetric QW layer structures to a different substrate. Furthermore, transferring to a transparent substrate enabled transmission measurements. For characterization, samples were transferred to sapphire because of its transparency to near-IR and visible wavelengths, its weak second order non-linearity, and compatibility with flip-chip substrate transfer procedures. Etch release layers of GaAs and Al$_{0.55}$Ga$_{0.45}$As were included in the epitaxial layer stack to facilitate transfer. The sapphire substrate was bonded to the epitaxial surface with epoxy, then mechanical lapping was used to etch the GaAs substrate, followed by removal of the remaining AlGaAs and GaAs etch-stop layers with a series of wet chemical etches. The growth terminated with a 10 nm GaAs cap layer, which prevented oxidation of the surface. After substrate transfer, the resulting flip-chip samples had a GaAs cap layer above the asymmetric QW layers, followed by the sapphire substrate. 


\subsection{Rotation-angle second harmonic generation experiments}

Surface second harmonic generation was used to excite SHG from the thin-film coupled QW samples \cite{foster_study_2022, bloembergen_light_1962, shen_surface_1986}. The measurements at 1550 nm fundamental wavelength were performed in transmission using the samples transferred to sapphire. The 1550 nm fundamental (excitation) was incident on the sample at a polar angle of 45$^o$ to the surface normal. The tunable-wavelength source was a Carbide CB3-20W (Light Conversion) femto-second laser pumping an optical parametric amplifier (OPA). The sample was rotated through 360$^o$ while under constant illumination. The input and output light were polarized at either p or s polarization. All four combinations of input/output polarization were measured for the rotation-angle SHG studies. A combination of long-pass, short-pass, and neutral-density filters isolated the fundamental wavelength at the input and the second harmonic at the output. The neutral density (ND) filter at the input had a 1.8 OD to attenuate the fundamental signal, with an additional ND filter at the detector with an OD 0.35 to avoid detector saturation. The long-pass filter filtered out the SH wavelength from the OPA output with an OD greater than 5 below 1100 nm. The short-pass filter with an OD of above 5 for wavelengths longer than 950 nm was utilized before detection to filter out the 1550 nm fundamental wavelength. The SH output was measured using an avalanche photodiode (APD440A Thorlabs) and SR860 (Stanford Research Systems) lock-in amplifier locked to the laser trigger signal. Due to the four-fold rotation-inversion symmetry of the crystal, we observe four distinct peaks in the SH response with 360$^o$ sample rotation. The intensity dependence of SH, Fig. \ref{fig: Setup and SHG Signature}c, was measured using p-in and p-out polarizations at the rotation angle with maximum SHG. Fig. \ref{fig: Setup and SHG Signature} shows the transmission measurement setup.

\subsection{Electron microscopy characterization}

Thin sample foils for scanning transmission electron microscopy (STEM) and energy dispersive spectroscopy (EDS) were prepared using focused ion beam (FIB) milling. These specimen were subsequently fine-polished using Ar milling in a nano-mill. STEM images were acquired using a double aberration-corrected JEOL ARM-300CF microscope operating at 300 kV. EDS maps were acquired using dual 100 mm$^{2}$ Si drift detectors (SDDs). Z-contrast high angle annular dark field (HAADF-STEM) imaging was performed with a probe convergence semi-angle of 25.7 mrad and an inner collection angle of 53 mrad. For constructing EDS maps, 100 scans (each with 10 microsecond pixel dwell time and varying step sizes) in the same area were summed. 

The EDS composition profiles of Ga and Al along the growth direction were modeled with linear piecewise functions. The modeling captured the measured composition gradients and reached the target compositions of Al$_{0.00}$Ga$_{1.00}$As in the QW's and Al$_{0.55}$Ga$_{0.45}$As in the tunneling barrier. The fitted composition profiles were then used to simulate the bound state wavefunctions and energy levels, which were used to calculate $\chi^{(2)}$.

\subsection{Extracting effective $\chi^{(2)}$ from second harmonic generation measurements}

The method for determining the effective $\chi^{(2)}$ from second harmonic generation measurements relies on Lorentz reciprocity of the propagation of fundamental and second harmonic waves through the material. We also apply the undepleted pump approximation. First, the electric field at the fundamental and second harmonic wavelengths were simulated for the entire sample to capture standing wave effects. Next, the accepted values of $\chi^{(2)}_{GaAs}$, $\chi^{(2)}_{AlGaAs}$, and the measured SH from the GaAs and AlGaAs controls were used to calculate $\alpha$, which is a parameter relating the material $\chi^{(2)}$ and the measured second harmonic. The value of the second harmonic on the left side of Eq. \ref{eqn: extract MQW chi(2)} is the average of the four second harmonic peaks in the rotation-angle SHG measurements. This value is from measurements of the P-in/P-out polarization with a 45$^o$ polar angle. The analysis assumes the same zincblende crystal symmetry for the multi QWs as GaAs and AlGaAs. The effective $\chi^{(2)}$ of the coupled QW region was then calculated using eqn. \ref{eqn: extract MQW chi(2)}.

\begin{multline}
    \frac{SH [mV]}{\alpha} = |(\chi^{(2)}_{GaAs}\iint_{GaAs}E_{p}^{(2\omega)}(E_{p}^{(\omega)})^2dv + \chi^{(2)}_{AlGaAs}\iint_{AlGaAs}E_{p}^{(2\omega)}(E_{p}^{(\omega)})^2dv \\+
    \frac{1}{2}\chi^{(2)}_{MQW}\iint_{MQW}E_{p}^{(2\omega)}(E_{p}^{(\omega)})^2dv)|^2
    \label{eqn: extract MQW chi(2)}
\end{multline}

In Eq. \ref{eqn: extract MQW chi(2)}, $\chi^{(2)}_{GaAs} = 377$ pm/V is the accepted value of the bulk GaAs susceptibility, $\chi^{(2)}_{AlGaAs}= 290$ pm/V is the accepted value of the bulk GaAs susceptibility, $\alpha$ is a free parameter relating material susceptibility and the measured SHG, $E_{p}^{(\omega)}$ is the simulated field profile at the fundamental frequency, $E_{p}^{(2\omega)}$ is the simulated field profile at the SH frequency, and the integrals are taken over the GaAs, AlGaAs, and coupled QW regions, respectively. $\chi^{(2)}_{MQW}$ is the effective second-order susceptibility for the MQW region. All the repeated periods of the asymmetric coupled QW structures are considered to be one material layer with $\chi^{(2)}_{MQW}$. 3 nm of the period barrier on either side of the coupled QW structures are included with the coupled QWs, as the coupled QW bound state wavefunctions go to zero within this thickness. The rest of the period barrier AlGaAs layers are considered part of the AlGaAs bulk. 

The factor of $\frac{1}{2}$ in the multi-QW term of Eq. \ref{eqn: extract MQW chi(2)} is a unique result of the interband nonlinearity from the coupled QWs. Ordinarily for a bulk material of this crystal symmetry, $\chi^{(2)}_{xzx}=\chi^{(2)}_{xxz}$, so the polarization can be written $P_x(2\omega) = ... + 2\chi^{(2)}_{xzx}E_z(\omega)E_x(\omega)+...$, where the ellipsis indicates additional terms. However, considering the QW structure and the combination of interband and intersubband transition dipole matrix elements, as per Eq. \ref{eqn: chi(2) starting equation}, the susceptibility tensor elements are no longer equal, $\chi^{(2)}_{xzx} \neq \chi^{(2)}_{xxz}$. Furthermore,

\begin{equation}
    \chi_{xzx}^{(2)} \sim \langle\psi_{e,2}|x|\psi_{hh,2}\rangle\langle\psi_{hh,2}|z|\psi_{hh,2}\rangle\langle\psi_{hh,2}|x|\psi_{e,2}\rangle \neq 0
    \label{chi2_xzx}
\end{equation}
\begin{equation}
    \chi_{xxz}^{(2)} \sim \langle\psi_{e,2}|x|\psi_{hh,2}\rangle\langle\psi_{hh,2}|x|\psi_{hh,2}\rangle\langle\psi_{hh,2}|z|\psi_{e,2}\rangle \simeq 0
    \label{chi2_xxz}
\end{equation}

where $\chi^{(2)}_{xxz}\simeq0$ because the second and third transition dipole matrix elements in Eq. \ref{chi2_xxz} are near zero. This results in $P_x(2\omega) = ... + \chi^{(2)}_{xzx}E_z(\omega)E_x(\omega)+...$ without the typical factor of 2 with the susceptibility tensor element. To account for this, we need to introduce a factor of $\frac{1}{2}$ to the multi-QW term of Eq. \ref{eqn: extract MQW chi(2)}.
\subsection{Black-box machine-learning optimization of asymmetric coupled QW structure}

The described two-well coupled QW system can be further generalized to access higher $\chi^{(2)}$ by freeing layer compositions and the addition of wells, with designs with $\chi^{(2)}>7$~nm/V at 1550 nm suggested, shown in Fig. \ref{fig: Design Space}. In systems with many free parameters, careful exploration of a subset of the full design space is required to discover solutions with enhanced $\chi^{(2)}$. Such exploration can be conducted in a principled manner using bandit / blackbox optimization techniques as in the work of Bartlett \textit{et al.} \cite{bartlett_simple_2019}. In the limit of a large number of free parameters, we recover the digital alloy design space.

\backmatter

\bmhead{Supplementary information}

Please refer to the supplementary information. 

Please refer to Journal-level guidance for any specific requirements.

\bmhead{Acknowledgements}

This research was supported by a Multidisciplinary University Research Initiative from the Air Force Office of Scientific Research (AFOSR MURI Award No. FA9550-22-1-0307) and partially supported by the National Science Foundation (DMREF-1629330).

\section{Declarations}



\subsection{Funding}
This research was supported by a Multidisciplinary University Research Initiative from the Air Force Office of Scientific Research (AFOSR MURI Award No. FA9550-22-1-0307) and partially supported by the National Science Foundation (DMREF-1629330).

\subsection{Conflict of interest/Competing interests}
The authors declare no competing interests.

\subsection{Ethics approval and consent to participate}
Not applicable

\subsection{Consent for publication}

\subsection{Data availability}
Data is available by reasonable request.

\subsection{Materials availability}

\subsection{Code availability}
Code is available by reasonable request

\subsection{Author contribution}
R.R., P.D., K.C.W., Q.M., T.H., A.M.S., M.W., S.R.B. completed simulations for sample design.
A.F.R., R.R., A.M.S., S.R.B. carried out crystal growth.
H.J., A.F.R., R.R., I.B., S.R.B. performed sample preparation.
R.R., M.B., Z.S., F.S., M.C.D., D.W., S.R.B. carried out optical measurements.
M.W., X.P. were responsible for structural measurements.
R.R., S.E., P.D., M.B. K.C.W., S.M., M.C., M.W., S.S., X.P. D.W., A.A., J.B.K., S.R.B. performed analysis.
R.R. wrote the paper with assistance from A.F.R., P.D., M.B., S.E., M.W., A.A., D.W., S.R.B.
R.R., J.B.K., S.R.B. were responsible for project conceptualization
R.R. and S.R.B. were responsible for project coordination. 

\subsection{Acknowledgments}
The authors would like to acknowledge Prof. Edward T. Yu for helpful discussions in relation to theory. 

\noindent






\begin{appendices}



\end{appendices}


\bibliography{references_v2_Feb25_2026}

\end{document}